\documentclass{iau} 
\usepackage{graphicx}

\title[color variation on fast rotating asteroids] %% give here short title %%
{Searching for color variation on fast rotating asteroids with simultaneous {\it V-J} observations}

\author[David Polishook \& Nicholas Moskovitz]   %% give here short author list %%
{David Polishook$^1$
 \and Nicholas Moskovitz$^2$}

\affiliation{$^1$Weizmann Institute of Science, \\ Rhovot 7610001,
Israel \\ email: {\tt david.polishook@weizmann.ac.il} \\[\affilskip]
$^2$Lowell Observatory, \\ Flagstaff 86001, AZ, USA \\email: {\tt nmosko@lowell.edu}}

\pubyear{2015}
\volume{318}  %% insert here IAU Symposium No.
\jname{Asteroids: New Observations, New Models}
\editors{Steven Chesley, Alessandro Morbidelli \& Robert Jedicke, eds.}
\begin{document}

\maketitle

\begin{abstract}
Boulders, rocks and regolith on fast rotating asteroids ($<2.5$ hours) are modeled to slide towards the
equator due to a strong centrifugal force and a low cohesion force. As a result, regions of fresh subsurface material can be exposed. Therefore, we searched for color variation on small and fast rotating asteroids. We describe 
a novel technique in which the asteroid is simultaneously observed in the visible and near-IR 
wavelength range. In this technique, brightness changes due to atmospheric extinction effects can be calibrated across the visible and near-IR images. We use {\it V}- and {\it J}-band filters since the distinction in color between weathered and unweathered surfaces on ordinary chondrite-like bodies is most prominent at these wavelengths and can reach $\sim25\%$.
To test our method, we observed 3 asteroids with Cerro Tololo's 1.3 m telescope. We find $\sim5\%$ variation of the mean {\it V-J} color, but do not find any clearly repeating color signature through multiple rotations. This suggests that no landslides 
occurred within the timescale of space weathering, or that Landslides occurred but the exposed 
patches are too small for the measurements' uncertainty.
\end{abstract}

\firstsection % if your document starts with a section,
              
\section{Introduction and motivation}

The ``rubble pile spin barrier" represents a limit to the spin rates of asteroids: bodies larger than $200-300$ meters do not rotate faster than $2.0-2.2$ hours (Fig. 1; \cite[Pravec et al. 2002]{Pravec_etal02}). Since asteroids are weak aggregate bodies with negligible tensile strength, bound together only by gravity (``rubble piles" , \cite[Richardson et al. 1998]{Richardson_etal98}), an asteroid that rotates faster than the spin barrier will self-disrupt and lose mass due to high centripetal force. The spin barrier is not fixed but a function of the body's density, shape and value of cohesive strength (\cite[Holsapple 2007]{Holsapple_2007}). Moreover, it was suggested that asteroids can oscillate around the spin barrier: they spin-up due to thermal torques, shed mass due to their weak structure, and spin-down to conserve angular momentum. This process, referred to as the YORP-cycle (\cite[Rubincam 2000]{Rubincam_2000}, \cite[Vokrouhlick{\'y} et al. 2015]{Vokrouhlicky_etal2015}), suggests that some asteroids with fast rotation rate, might have crossed the spin barrier and suffered from mass loss. \cite[Walsh et al. (2008)]{Walsh_etal2008} used this model to show how fast rotation transports material towards the equator, forms a ridge (e.g. asteroid $1999 KW_4$; \cite[Harris et al. 2009]{Harris_etal2009}) and is then ejected into space. As a result, subsurface material should be exposed where slides occurred.

\begin{figure}[b]
% \vspace*{-2.0 cm}
\begin{center}
 \includegraphics[width=3.4in]{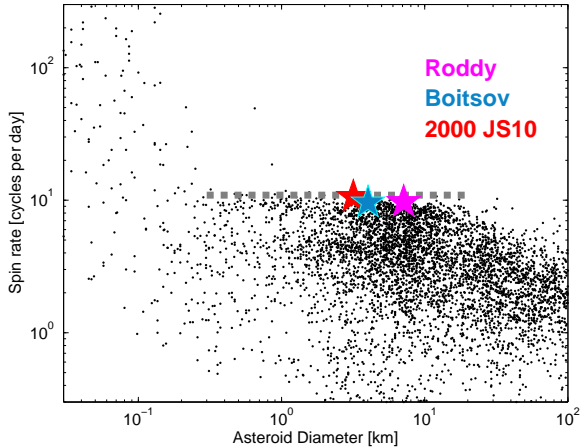} 
% \vspace*{-1.0 cm}
 \caption{The ``rubble pile spin barrier" (dashed-line), marked on the diameter-spin rate diagram (dots), with our three asteroid targets (stars).}
   \label{fig1}
\end{center}
\end{figure}

The surface of asteroids is constantly being impacted by solar wind particles, cosmic rays, and micrometeorites. These effects alter the regolith layer on planetary surfaces. For example, ordinary chondrite, i.e. S-type asteroids, display ``weathered" surfaces that are darker and have redder colors than their unweathered counterparts taxonomically known as Q-types (\cite[Clark et al. 2002]{Clark_etal2002}, \cite[DeMeo et al. 2009]{DeMeo_etal2009}). Asteroids with fresh surfaces were observed within the near-Earth population (\cite[Binzel et al. 2010]{Binzel_etal2010}), and among asteroids disintegrated due to fast rotations (``asteroid pairs"; \cite[Polishook et al. 2014a]{Polishook_etal2014a}). The timescale of space weathering ranges between $\sim10^5$ years (\cite[Nesvorn{\'y} et al. 2010]{Nesvorny_etal2010}) to $\sim10^6$ years (\cite[Vernazza et al. 2009]{Vernazza_etal2009}), is shorter than typical collisional ages of asteroids (\cite[Bottke et al. 2005]{Bottke_etal2005}, \cite[Marzari et al. 2011]{Marzari_etal2011}), thus all asteroids should present ``weathered'' colors unless mass shedding happened recently. In the case of S-type asteroids, we predict that they will present Q-type colors, at least partly, at the locations of disruption. Finding an asteroid with both Q- and S-type colors would support the model of asteroids oscillating around the ``rubble pile spin barrier'' and would confirm that these two taxonomic types are mineralogically equal.

\section{Method and Observations}

\begin{figure}[b]
% \vspace*{-2.0 cm}
\begin{center}
 \includegraphics[width=3.4in]{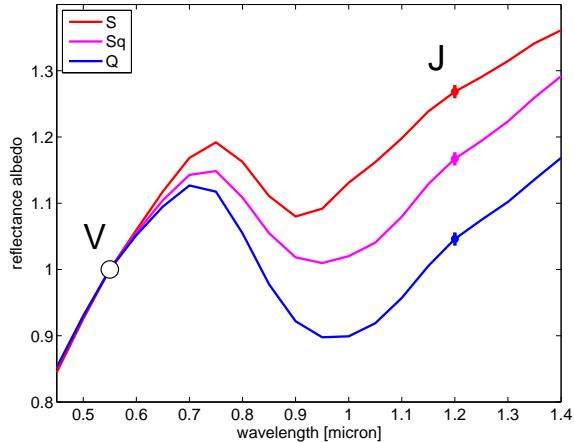} 
% \vspace*{-1.0 cm}
 \caption{An average weathered (red), intermediate (magenta) and fresh (blue) reflectance spectra representing the S-, Sq- and Q-type taxonomies, respectively (DeMeo et al. 2009). At the J-band the three curves are the most distinctive with $\sim25\%$ difference in spectral slope.}
   \label{fig2}
\end{center}
\end{figure}

Detecting color variation on small and fast rotating asteroids is difficult with spectroscopy due to the influence of systematic errors on spectral slope and the need for short exposure times to isolate discrete rotational phases of the body. Broadband photometry is also problematic since it introduces large systematic errors when images in different filters are calibrated with standard stars (see the dispute about color variation on asteroid (832) Karin: \cite[Sasaki et al. 2004]{Sasaki_etal2004}, \cite[Chapman et al. 2007]{Chapman_etal2007}, \cite[Vernazza et al. 2007]{Vernazza_etal2007}). Therefore, we employed a novel technique in which the asteroid is simultaneously observed in the visible ($0.4-0.8~\mu m$) and near-IR ($0.8-2.5~\mu m$) wavelength range using an instrument that employs a dichroic and separate near-IR and visible detectors. In this way atmospheric extinction effects can be correct between the visible and the near-IR images. We use {\it V}- and {\it J}-band filters since the distinction between fresh and weathered surfaces are most prominent in these wavelengths and reach $\sim25\%$ (Fig.\,\ref{fig2}).

Our asteroid sample includes three fast rotating asteroids ($P<2.5$ hours) just below the ``rubble pile spin barrier" (Fig.\,\ref{fig1}): (3873) Roddy, (6685) Boitsov and (30311) 2000 JS10. The targets' diameters range between 3 and 7 km, which makes them sensitive to the YORP effect (\cite[Polishook \& Brosch 2009]{PolishookBrosch2009}, \cite[Marzari et al. 2011]{Marzari_etal2011}). The lightcurve amplitude of all three asteroids is of order 0.1 magnitude, suggesting their shape is not elongated but rather spherical, similar to the shape of asteroid 1999 KW4 and the asteroids modeled by \cite[Walsh et al. (2008)]{Walsh_etal2008} model.

Observations took place using Cerro Tololo's 1.3 m telescope and the ANDICAM detector on April 2014. As described above, this detector includes a dichroic and the visible and infrared beams are recorded simultaneously. Exposure times of 90 seconds were used to balance between minimizing the sampling interval and maximizing the SNR. The asteroids were observed during 2 rotational cycles to confirm features on the color-curve. See Table I for observational details. Standard reduction process included bias subtraction and flat field correction. Measurements were done with an aperture of 5 pixels. We calculate the flux ratio between every visible and near-IR point that were taken within 90 seconds of one another.

\begin{table}
  \begin{center}
  \caption{Summary of observations$^1$}
  \label{tab1}
 {\scriptsize
%  \begin{tabular}{|l|c|c|c|c|clclcl}\hline 
  \begin{tabular}{cccccccccc}\hline 
{\bf Asteroid} & {\bf Diameter} & {\bf Period} & {\bf Date} & {\bf Filter} & {\bf Time Span} & {\bf Images Num.} &  {\bf r} & {\bf $\Delta$} & {\bf $\alpha$} \\
          & {\bf [km]} & {\bf [hours]} & & & {\bf [hours]}  & & {\bf [AU]} & {\bf [AU]} & {\bf [degrees]} \\ \hline
3873 & 7.13 & 2.4782 & 20140415 & V & 2.45 & 23 & 1.7 & 0.8 & 23.1 \\ 
          & & & 20140415 & J & 2.52 & 38 & 1.7 & 0.8 & 23.1 \\ 
          & & & 20140416 & V & 3.19 & 78 & 1.7 & 0.8 & 23.1 \\ 
          & & & 20140416 & J & 3.19 & 91 & 1.7 & 0.8 & 23.1 \\ \hline
6685 & 3.6 & 2.5 & 20140405 & V & 2.65 & 52 & 2.2 & 1.2 & 3.8 \\
          & & & 20140405 & J & 2.65 & 71 & 2.2 & 1.2 & 3.8 \\
          & & & 20140406 & V & 2.73 & 58 & 2.2 & 1.2 & 4.4 \\ 
          & & & 20140406 & J & 2.73 & 83 & 2.2 & 1.2 & 4.4 \\ \hline
30311 & 3.67 & 2.266 & 20140412 & V & 2.69 & 64 & 1.9 & 1.0 & 20.4 \\
          & & & 20140412 & J & 2.69 & 76 & 1.9 & 1.0 & 20.4 \\
          & & & 20140413 & V & 2.69 & 69 & 1.9 & 1.0 & 19.9 \\ 
          & & & 20140413 & J & 2.68 & 82 & 1.9 & 1.0 & 19.9 \\

  \end{tabular}
  }
 \end{center}
\vspace{1mm}
 \scriptsize{
  $^1$Legend: Asteroid name, observation date, filter, nightly time span, images number, heliocentric and geocentric distance, phase angle. \\
}
\end{table}

\begin{figure}[b]
% \vspace*{-2.0 cm}
\begin{center}
 \includegraphics[width=3.4in]{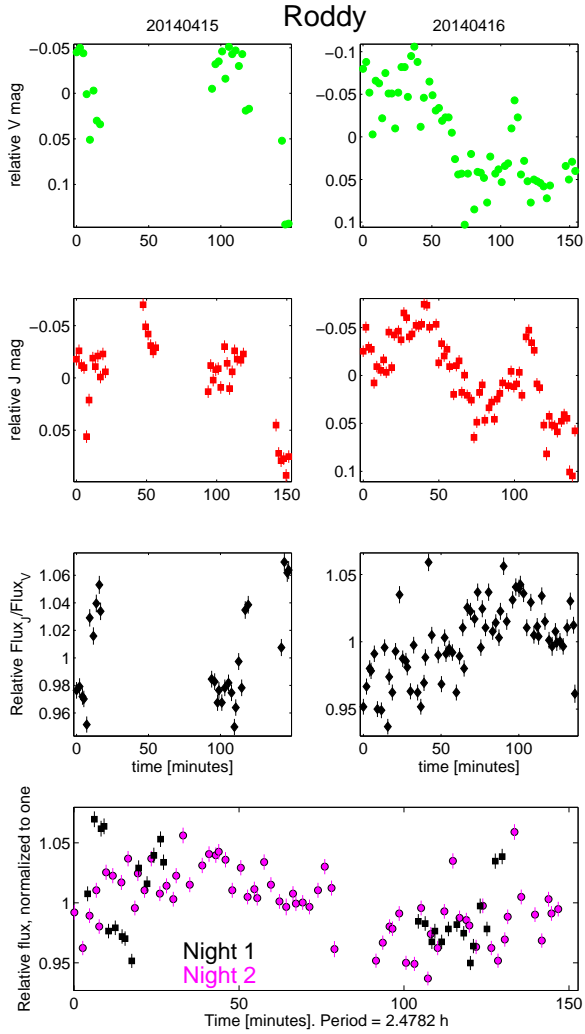} 
% \vspace*{-1.0 cm}
 \caption{Photometric measurements for asteroid Roddy. Top: {\it V}-band lightcurves measured on April 15 \& 16. Second row: {\it J}-band lightcurves. Third row: relative fluxes in {\it J} and {\it V}. Bottom row: the curves of relative fluxes from the two nights are folded to a single rotation cycle. The Y-axis is normalized to one to allow the comparison between the two nights.}
   \label{fig3}
\end{center}
\end{figure}

\begin{figure}[b]
% \vspace*{-2.0 cm}
\begin{center}
 \includegraphics[width=3.4in]{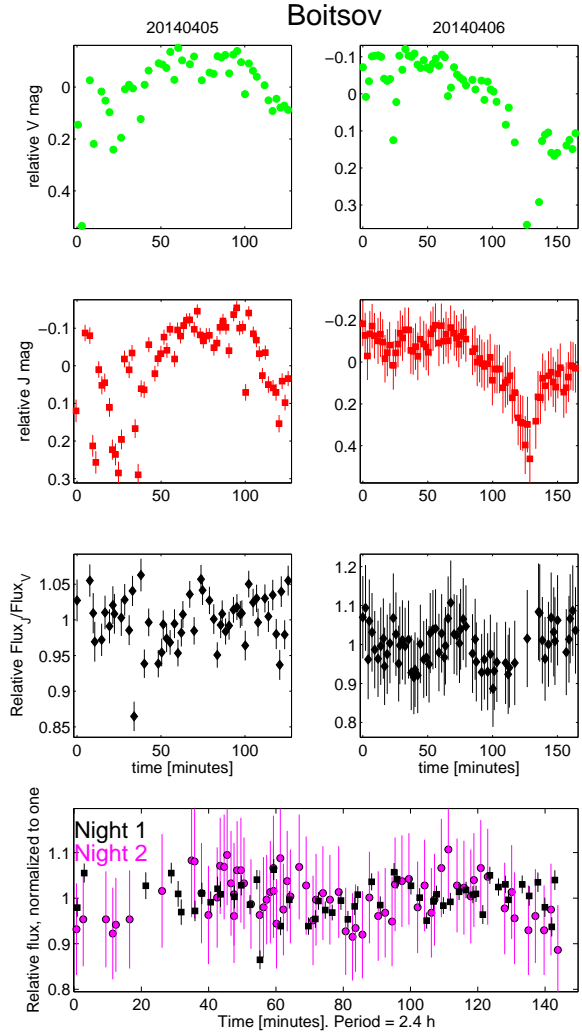} 
% \vspace*{-1.0 cm}
 \caption{Such as Fig.\,\ref{fig3} for asteroid Boitsov.}
   \label{fig4}
\end{center}
\end{figure}

\begin{figure}[b]
% \vspace*{-2.0 cm}
\begin{center}
 \includegraphics[width=3.4in]{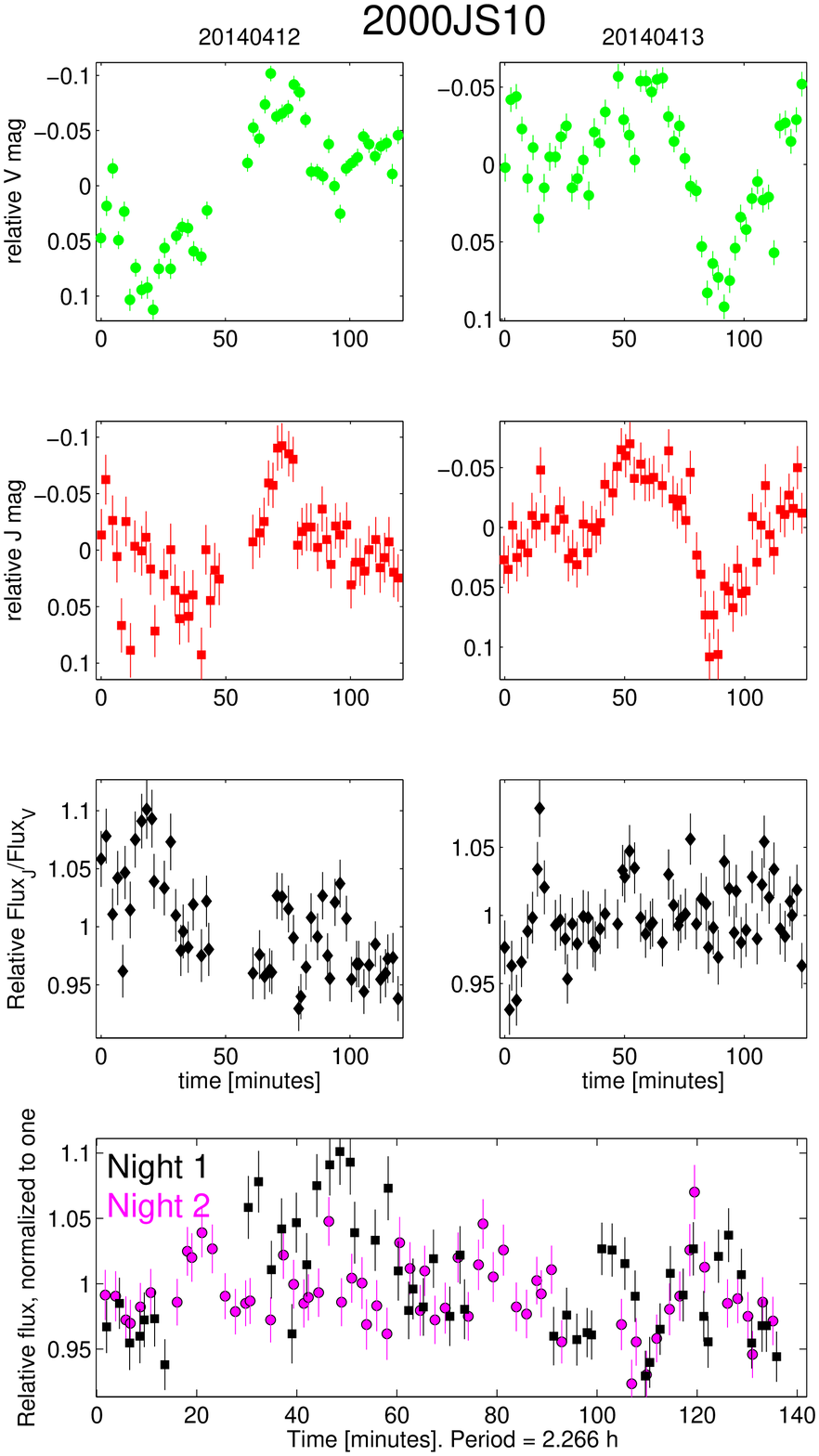} 
% \vspace*{-1.0 cm}
 \caption{Such as Fig.\,\ref{fig3} for asteroid 2000JS10.}
   \label{fig5}
\end{center}
\end{figure}

\section{Results and conclusions}

The lightcurves in the {\it V}- and {\it J}-bands and the relative flux between the measurements are presented in Fig.\,\ref{fig3}-\,\ref{fig5}. The different patterns in the lightcurves are due to atmospheric disturbances that affect the measurements in both {\it V}- and {\it J}-band in the same way. Unfortunately, from the lightcurves it is noticeable that the atmospheric conditions were not photometric (especially on April 5 and 6 where the magnitude dropped by half a magnitude), and even though most of the disturbances are calibrated out it adds to the overall scatter of the flux ratio. The $J/V$ flux ratio is approximately $\pm5\%$ for most measurements. In two extreme cases a scatter of $\pm10\%$ is measured (for Boitsov on April 6, and 2000JS10 on April 12). However, this variability is not repeated on both nights and it is likely due to systematic errors rather than a real difference in surface reflectivity. While some continuous variation in the flux ratio is noticeable (e.g., Boitsov on April 6, 2000JS10 on Apr 12) it does not repeat in the other night, therefore, we cannot confirm it as real variation. A $\sim10\%$ variation measured on the flux ratio of Roddy on April 16, cannot be confirmed or rejected since the asteroid was only partly observed on the other night (April 15). Since the expected flux ratio between S- and Q-type asteroids for the {\it J/V} flux ratio is $\sim25\%$, we can determine we did not detect Q-type, fresh colors on the surface.

The results suggest one of the following conclusions:

1. The asteroids never reached a spin rate beyond the 'rubble pile spin barrier', therefore, no mass loss occurred. 

2. No landslides occurred within the timescale of space weathering. Since the timescale of space weathering is estimated as $10^5$ to $10^6$ years, the three asteroids might have experienced slides longer than this time limit. This makes sense since the YORP effect timescale for asteroids as large as the three targets is in order of a few million years (\cite[Marzari et al. 2011]{Marzari_etal2011}, \cite[Durech et al. 2012]{Durech_etal2012}, \cite[Rozitis et al. 2013]{Rozitis_etal2013}). Therefore, even if slides happened, the fresh area is weathered by now. If true, examining other fast rotating asteroids can be used in order to constrain the time of their possible slides events or disintegrations.

3. Landslides occurred but the exposed patches are too small for the measurementsÕ uncertainty. If true, then evolution across the rubble pile spin barrier only involves a small amount of mass loss or reshaping of the body. In order to test this possibility, data with higher signal-to-noise are required.

4. Slides occurred but did not expose fresh material. This statement contradicts the understanding that space weathering occurs only in nano- to micrometer-scale depths of an atmosphere-less body (\cite[Clark et al. 2002]{Clark_etal2002}, \cite[Noguchi et al. 2011]{Noguchi_etal2011}), while the slides occur on larger scale from micro- to kilo-meters (\cite[Scheeres 2007]{Scheeres_2007}, \cite[Walsh et al. 2008]{Walsh_etal2008}, \cite[Pravec et al. 2010]{Pravec_etal2010}, \cite[Sanchez \& Scheeres 2014]{SanchezScheeres2014}). This would be consistent with the claim that landslides or disintegration events are followed by ejection of a dust coma, and that this dust can compositionally homogenize the asteroidÕs surface as it is re-accreeted (\cite[Jewitt et al. 2014]{Jewitt_etal2014}, \cite[Polishook et al. 2014b]{Polishook_etal2014b}). If true, this means that color variation on asteroid surfaces will be muted in cases of rotational disintegration.

To conclude, we describe a simple and novel method to search for color variation on asteroid surfaces. Even though we could not find color variation on the three observed targets, employing this observing technique on larger telescopes, while observing smaller asteroids that rotate faster, could reveal surface color heterogeneity and thus a means to contain the evolutionary history of asterods.

\end{document}